\documentclass[manuscript,times]{derivative_template}

\usepackage{amsmath,amssymb,amsfonts,amsthm}%

\usepackage{journal_abbreviations}
\usepackage[dvipsnames]{xcolor}

\makeatletter
\PassOptionsToPackage{numbers,square,sort&compress}{natbib}% if not yet loaded
\@ifpackageloaded{natbib}{%
\def\NAT@cmprs{1}%
  \setcitestyle{numbers,square,comma,sort&compress}% public API
  \bibpunct{[}{]}{,}{n}{}{,}% numeric mode + square brackets
}{%
  \AtBeginDocument{%
    \setcitestyle{numbers,square,comma,sort&compress}%
    \bibpunct{[}{]}{,}{n}{}{,}%
  }%
}
\makeatother

\makeatletter
\renewcommand{\frontmatter@title@format}{%
  \centering
  \bfseries\LARGE
  \linespread{1.05}\selectfont         % relaxed line spacing
  \setlength{\emergencystretch}{2em}  % help LaTeX justify nicely
}
\makeatother

\begin{document}

\title{\Large The Fate of Hydrogen and Helium: From Planetary Embryos to Earth- and Neptune-like Worlds\vspace{2mm}}

\correspondingauthor{Akash Gupta, Jie Deng}
\email{akashgpt@princeton.edu; jie.deng@princeton.edu}

\author[0000-0002-2006-7769]{Akash Gupta}
% \altaffiliation{51 Pegasi b Fellow}
% \altaffiliation{Future Faculty in Physical Sciences Fellow}
% \altaffiliation{Harry H. Hess Postdoctoral Fellow}
\affiliation{Department of Astrophysical Sciences, Princeton University, Princeton, NJ 08550, USA}
\affiliation{Department of Geosciences, Princeton University, Princeton, NJ 08550, USA}

\author[0000-0003-2587-3185]{Haiyang Luo}
\affiliation{Department of Geosciences, Princeton University, Princeton, NJ 08550, USA}

\author[0000-0001-5441-2797]{Jie Deng}
\affiliation{Department of Geosciences, Princeton University, Princeton, NJ 08550, USA}

\author[0000-0002-3099-5024]{Adam Burrows}
\affiliation{Department of Astrophysical Sciences, Princeton University, Princeton, NJ 08550, USA}

\begin{abstract}
Hydrogen, helium, silicates, and iron are key building blocks of rocky and gas-rich planets, yet their chemical interactions remain poorly constrained. Using first-principles molecular dynamics and thermodynamic integration, we quantify hydrogen and helium partitioning between molten silicate mantles and metallic cores for Earth-to-Neptune-mass planets. Hydrogen becomes strongly siderophilic above $\sim$25 GPa but weakens beyond $\sim$200 GPa, whereas helium remains lithophilic yet increasingly soluble in metal with pressure. Incorporating these trends into coupled structure–chemistry models suggests that majority of hydrogen and helium reside in planetary interiors, not atmospheres, with abundances strongly depending on planet mass. Such volatile exchange may influence the redox states of secondary atmospheres, longevity of primordial envelopes, predicted CHNOPS abundances, and emergence of helium-enriched atmospheres, while He 1083 nm and H Lyman-$\alpha$ lines provide potential probes of atmosphere-interior exchange. These findings link atomic-scale interactions to planetary-scale observables, providing new constraints on the origins of Earth-to-Neptune-sized worlds.
\end{abstract}

% \keywords{exoplanets, terrestrial planets, planet atmospheres, planet interiors, atomistic simulations, statistical mechanics, hydrogen, helium, volatile partitioning, atmosphere-interior interaction, interdisciplinary astronomy}

% .\phantom{.}\vspace{100mm}

% \newpage
\section*{\phantom{Main Text}}%\vspace{10mm}}
% \section*{Main Text}%\vspace{10mm}}
% \newpage

Hydrogen (H), helium (He), iron (Fe), and silicates (e.g., MgSiO$_3$) are the primary building blocks of planets. These materials dominate planetary atmospheres and interiors, and together shape the origin, evolution, structural and compositional diversity of Earth- and Neptune-like planets. Yet, despite their central role in shaping such worlds and interpreting atmospheric observations, the degree to which these materials chemically interact remains poorly understood.

H and He---two most abundant elements in the Universe---comprise the bulk of the protoplanetary disks in which planets form. Evidence for this comes from exoplanet population studies of the radius valley~\citep{fulton2017a,lopez2013a,owen2013a,owen2017a,ginzburg2018a,gupta2020a}, the gas-rich planets in the Solar System~\citep{guillot2015a,helled2020a}, isotopic anomalies in Earth and meteorites~\citep{mukhopadhyay2012a,mukhopadhyay2019a}, and direct observations of accreting disks~\citep{keppler2018a,muller2018a,vanCapelleveen2025a}. These lines of evidence collectively indicate that the typical planet accretes an optically thick, H/He-dominated envelope during formation. While some planets later lose these primordial gases to become rocky “super-Earths”, others retain them as sub-Neptunes. Because such H/He atmospheres possess large scale heights, they are especially amenable to spectroscopic characterization, making them prime targets for James Webb Space Telescope (JWST) and future observatories such as Ariel, the ELTs, and the Habitable Worlds Observatory~\citep{radica2024a,benneke2024a,deWit2025a}. Observations of escaping H and He via the H Lyman-$\alpha$ and He 1083 nm lines confirm the role of atmospheric loss in sculpting this population and further underscore the significance of these species~\citep{ehrenreich2012a,dossantos2023a,zhang2023a}.

While H and He dominate the outer envelopes, Fe and silicates form the bulk of planetary interiors and chondritic building blocks---roughly one-third and two-thirds of Earth’s mass, respectively~\citep{wang2008a}. Most super-Earths and sub-Neptunes are consistent with Earth-like silicate mantles and Fe-rich metallic cores comprising more than 90\% of total mass~\citep{rogers2015a,gupta2019a,rogersj2020a}, though water-rich variants have been proposed, particularly around M dwarfs~\citep{zeng2019a,madhusudhan2021a,luque2022a}.

Despite how fundamental these materials are to the concept of a “planet,” most models assume minimal chemical interaction between the H-He-dominated atmosphere and the Fe- and silicate-rich interior. This notion is inaccurate and could bias inferences of atmosphere-interior composition, mass–radius relations, thermal evolution, and atmospheric loss. Emerging theoretical and experimental work increasingly challenges this paradigm~\citep{chachan2018a,kite2020a,tagawa2021a,schlichting2022a,markham2022a,young2023a,luo2024a,gupta2025a,bower2025a,lichtenberg2025a,nixon2025a}. However, the relevant conditions deep inside planetary envelopes and interiors involve extreme pressures and temperatures that often exceed current laboratory capabilities. Constraints on the relevant phase equilibria are thus limited~\citep{luo2024a,gupta2025a}, forcing extrapolations over orders of magnitude.

Here, we investigate how H and He partition between silicate and metallic melts, or equivalently, between a silicate magma ocean and a liquid metallic core, across pressures and temperatures relevant to the formation and evolution of Earth- to Neptune-sized planets. Using \textit{ab initio} calculations benchmarked against laboratory and computational data, and a coupled planetary chemical-equilibrium–interior-structure model, we quantify the equilibrium compositions of atmospheres, mantles, and cores. We use these results to examine how volatile partitioning influences planetary origin and evolution, including mass–radius relations, atmospheric escape, composition---including {CHNOPS} species---and the diagnostic potential of He as a tracer of atmosphere–interior exchange.

\vspace{3mm}
\section*{{Metal--silicate partitioning of hydrogen and helium}}

To quantify the degree to which H and He dissolve in silicate and metallic melts, we perform \textit{ab initio} molecular dynamics (AIMD) simulations with thermodynamic integration (TI). For clarity and computational tractability, end-member configurations of molten MgSiO$_3$ and Fe are assumed to be representative of silicate and metallic melts, respectively. In turn, these melts are taken as analogs for a planet’s global molten reservoirs---a silicate magma ocean and a liquid metallic core. While the precise nature of interaction will vary in more complex natural melts, we expect the fundamental trends of H and He's siderophilicity or lithophilicity to be robust~\citep{luo2024b}.

TI-AIMD provides a rigorous framework to determine free energies of mixtures of complex, interacting systems, such as a system of H mixed with silicates or iron. In this approach, the system of interest is gradually transformed from an analytically tractable classical reference system, such as the ideal gas model, to a fully interacting system within, for instance, the density functional theory framework, by coupling them through a $\lambda$-dependent Hamiltonian. The free energy difference is then obtained by integrating the ensemble-averaged energy derivatives along this reversible pathway. This procedure circumvents the absence of methods to directly compute the free energy of a system, while retaining DFT-level accuracy in the treatment of electronic structure and ionic interactions. TI-AIMD thus yields internally consistent chemical potentials, which, when compared for the iron-rich compositions and silicate-rich compositions, allow quantitative predictions of partitioning or solubilities and phase equilibria inaccessible to experiment or static approximations (see \textit{Methods})~\citep{luo2024a,dorner2018a,li2020a,li2022a,zhang2012a,xiong2021a}. In addition to TI-AIMD simulations, we also perform two-phase \textit{ab initio} molecular dynamics (2P-AIMD) simulations of H and He in coexisting metal-silicate phases to gain further insights into their lithophilic or siderophilic nature.

\begin{figure}
\centering
    \includegraphics[width=\textwidth,trim=0 0 0 0,clip]{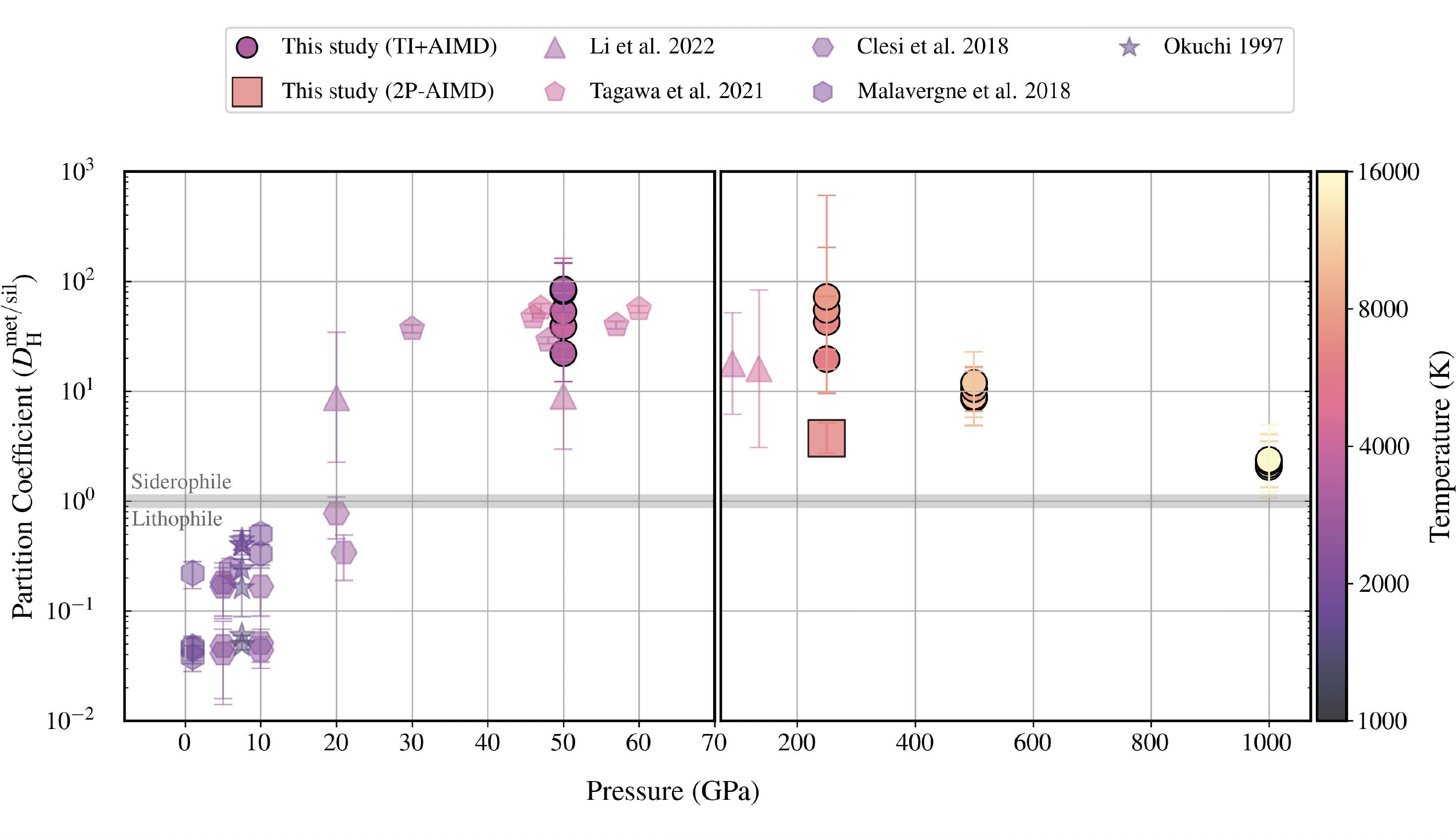}
    \caption{Partition coefficient for hydrogen between silicate and metallic melts, $D_{\mathrm{H}}^\mathrm{met/sil}$ as a function of temperature and pressure. Each point represents a unique $\{P, T\}$ condition. Circles denote thermodynamic-integration \textit{ab initio} molecular dynamics (TI-AIMD) results from this study with $x_{\mathrm{H}}^{\mathrm{sil}} \lesssim 0.2$, while the square indicates a two-phase AIMD (2P-AIMD) calculation at $x_{\mathrm{H}}^{\mathrm{sil}} \sim 0.42$. Other symbols show experimental~\citep{tagawa2021a,malavergne2019a,clesi2018a,okuchi1997a} and computational data~\citep{li2020a} from previous work. The horizontal gray line separates the siderophile (iron-loving) and lithophile (rock-loving) regimes, illustrating that hydrogen's strongly siderophilic nature at high temperatures and pressures.
    } 
     \label{fig:partition_coeff__H}
\end{figure}

\begin{figure}
\centering
    \includegraphics[width=\textwidth,trim=0 0 0 0,clip]{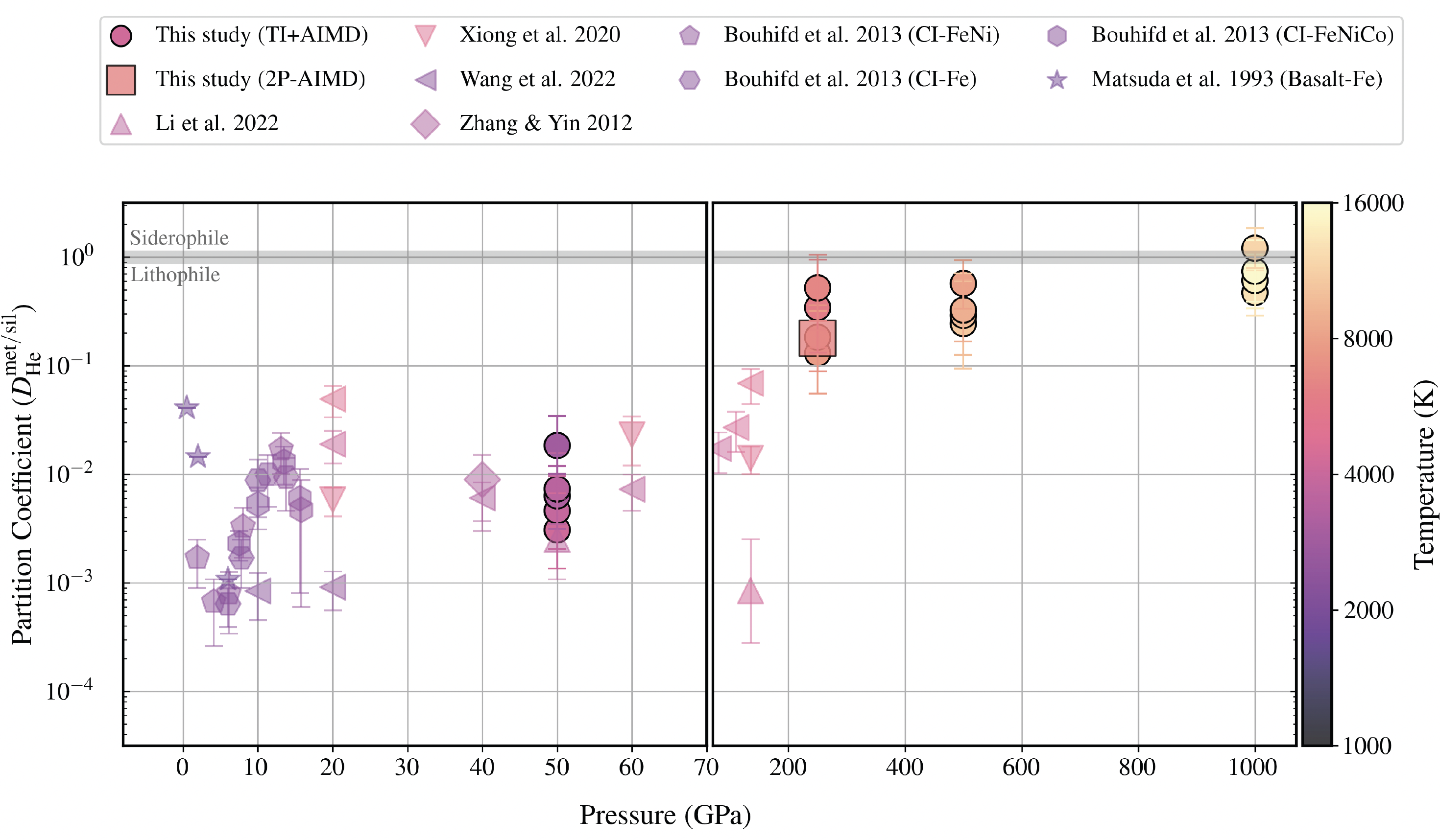}
        \caption{Partition coefficient for helium between silicate and metallic melts, $D_{\mathrm{He}}^\mathrm{met/sil}$ as a function of temperature and pressure. Each point represents a unique $\{P, T\}$ condition. As in~Figure~\ref{fig:partition_coeff__H}, circles denote thermodynamic-integration \textit{ab initio} molecular dynamics (TI-AIMD) results from this study with $x_{\mathrm{He}}^{\mathrm{sil}} \lesssim 0.2$, while the square indicates a two-phase AIMD (2P-AIMD) calculation at $x_{\mathrm{He}}^{\mathrm{sil}} \sim 0.49$. Other symbols show experimental~\citep{bouhifd2013a,matsuda1993a} and computational data~\citep{li2022a,zhang2012a,xiong2021a,wang2022a} from previous work. The horizontal gray line separates the siderophile (iron-loving) and lithophile (rock-loving) regimes, demonstrating that He remains largely lithophilic, with an increasing affinity towards iron at higher pressures.
    } 
     \label{fig:partition_coeff__He}
\end{figure}

To quantify the partitioning of H and He in the silicate (`sil') and metallic (`met') phases, i.e. for reactions
\begin{align}
\mathrm{H}_{\text{sil}}   &\;\rightleftharpoons\;   \mathrm{H}_{\text{met}} \quad \text{and}\\
\mathrm{He}_{\text{sil}}  &\;\rightleftharpoons\;   \mathrm{He}_{\text{met}},
\end{align}
we define ``partition coefficient" $D_{i}^\mathrm{met/sil}$ for species $i \in \{\mathrm{H},\,\mathrm{He}\}$: 
\begin{equation}
D_{i}^{\mathrm{met/sil}} = 
\frac{w_{i}^{\mathrm{met}}}{w_{i}^{\mathrm{sil}}},
\label{eq:D_w_defn}
\end{equation}
where $w_{i}^{j}$ is the weight fraction of species $i$ in phase $j \in \{\mathrm{metallic}, \mathrm{silicate}\}$. Equivalently, one can define an exchange coefficient  $K_{\text{D},\;i}^{\text{met/sil}} = {x_{i}^{\mathrm{met}}}/{x_{i}^{\mathrm{sil}}}$, where  $x_{i}^{j}$ is the corresponding mole fraction. For the purpose of discussion, we focus on $D_{i}^\mathrm{met/sil}$, or simply, $D$. 

With TI-AIMD simulations, we estimate free energies of mixtures of varying concentrations of H and He in MgSiO$_3$ and Fe, which then allows us to estimate $D_{i}^{\text{met/sil}}$ using principles from statistical thermodynamics (see \textit{Methods}). These calculations are performed across a vast range of temperatures and pressures relevant to Earth- to Neptune-like planets: $\sim$2,800-4,200 K at 50 GPa, 5,600-7,800 K at 250 GPa, 8,000-12,000 K at 500 GPa, and 12,000-16,000 K at 1,000 GPa. Past experimental studies only go up to 60 GPa for H~\citep{tagawa2021a,malavergne2019a,clesi2018a,okuchi1997a} and 15 GPa for He~\citep{bouhifd2013a,matsuda1993a}, and there are only one to two past computational studies for H and He that reach pressures of up to 135 GPa~\citep{li2020a,li2022a,zhang2012a,xiong2021a,wang2022a}. While the core-mantle boundary (CMB) pressure is $\sim$20 GPa for a Mars-mass embryo, it is $\sim$135 GPa for Earth and $\sim$1,000 GPa for a 10 Earth mass super-Earth or sub-Neptune such as K2-18 b (assuming Earth-like interior composition).

Our partition coefficients are consistent with both the magnitudes and trends reported previously (Figures~\ref{fig:partition_coeff__H} and~\ref{fig:partition_coeff__He}). In particular, for hydrogen, the values at 50 GPa and 3500 K are in excellent agreement with the high-pressure experiments of Tagawa et al.~\citep{tagawa2021a} and fall within 1-$\sigma$ of the TI-AIMD results of Li et al.~\citep{li2020a}. For He, our 50 GPa results likewise align well with prior computational studies, including Zhang et al., Xiong et al., Wang et al. and Li et al.~\citep{li2022a,zhang2012a,xiong2021a,wang2022a}. 

Furthermore, for both H and He, our 2P-AIMD simulations at 250 GPa and 6500 K also yield partitioning behaviors consistent with the TI-AIMD results---siderophilic for H and lithophilic for He. Quantitatively, He partition coefficients agree within 1-$\sigma$, whereas H values deviate by up to 2-$\sigma$, reflecting both the suppression of H partitioning by O and Si atoms and the finite-size effects that necessitate higher solute concentrations in 2P-AIMD simulations ($x_{\text{H or He}}^{\mathrm{sil}} \sim 0.4$–0.5) compared to TI-AIMD calculations ($x_{\text{H or He}}^{\mathrm{sil}} \lesssim 0.2$).

For both H and He, the partition coefficient $D$ varies systematically with pressure, temperature, and composition (Figures~\ref{fig:partition_coeff__H} and~\ref{fig:partition_coeff__He}). Above $\sim$50 GPa, pressure exerts the strongest influence, driving changes of 1–3 orders of magnitude over several hundred gigapascals. By contrast, the temperature dependence weakens at high pressure: at $T \gtrsim 3000$–4000 K, variations are limited to factors of a few over thousands of kelvin. Dependence on composition is negligible, except when $x_\mathrm{H} \gtrsim 0.2$.

Although both species exhibit qualitatively similar dependencies on $P$, $T$, and $x_i$, their trends diverge in detail. For H, $D$ rises steeply at low pressures, increasing by 3–4 orders of magnitude and reaching siderophilic values at core–mantle boundary conditions relevant to Mars, $\sim$20 GPa. Between $\sim$50 and 200 GPa, $D$ approaches values near $10^2 - 10^3$ at 1 $\sigma$ before declining gradually; H nevertheless remains a siderophile up to 1000 GPa, similar to core–mantle boundary conditions relevant to K2-18 b like sub-Neptunes, where $D \sim$ 2–3.

On the other hand, for He, $D$ varies rapidly at pressures $\lesssim$ 20 GPa although the reported values stay within $\sim$10$^{-3}$-10$^{-2}$, saturating to $\sim$10$^{-2\,\pm0.5}$ until $\sim$70 GPa. Overall, the partition coefficient tends to weakly increase with pressure. This trend is more pronounced at higher pressures of $\gtrsim$70 GPa, with $D$ reaching nearly 0.5-1 at 1,000 GPa. Thus, across the pressure range of 0.5 to 1000 GPa, He remains a lithophile, though its affinity for silicates diminishes as pressure increases.

\vspace{3mm}
\section*{{Implications of atmosphere--interior coupling across planetary masses}}\label{sec:discussion}

\begin{figure}
\centering
    \includegraphics[width=1\textwidth,trim=50 0 80 -0,clip]{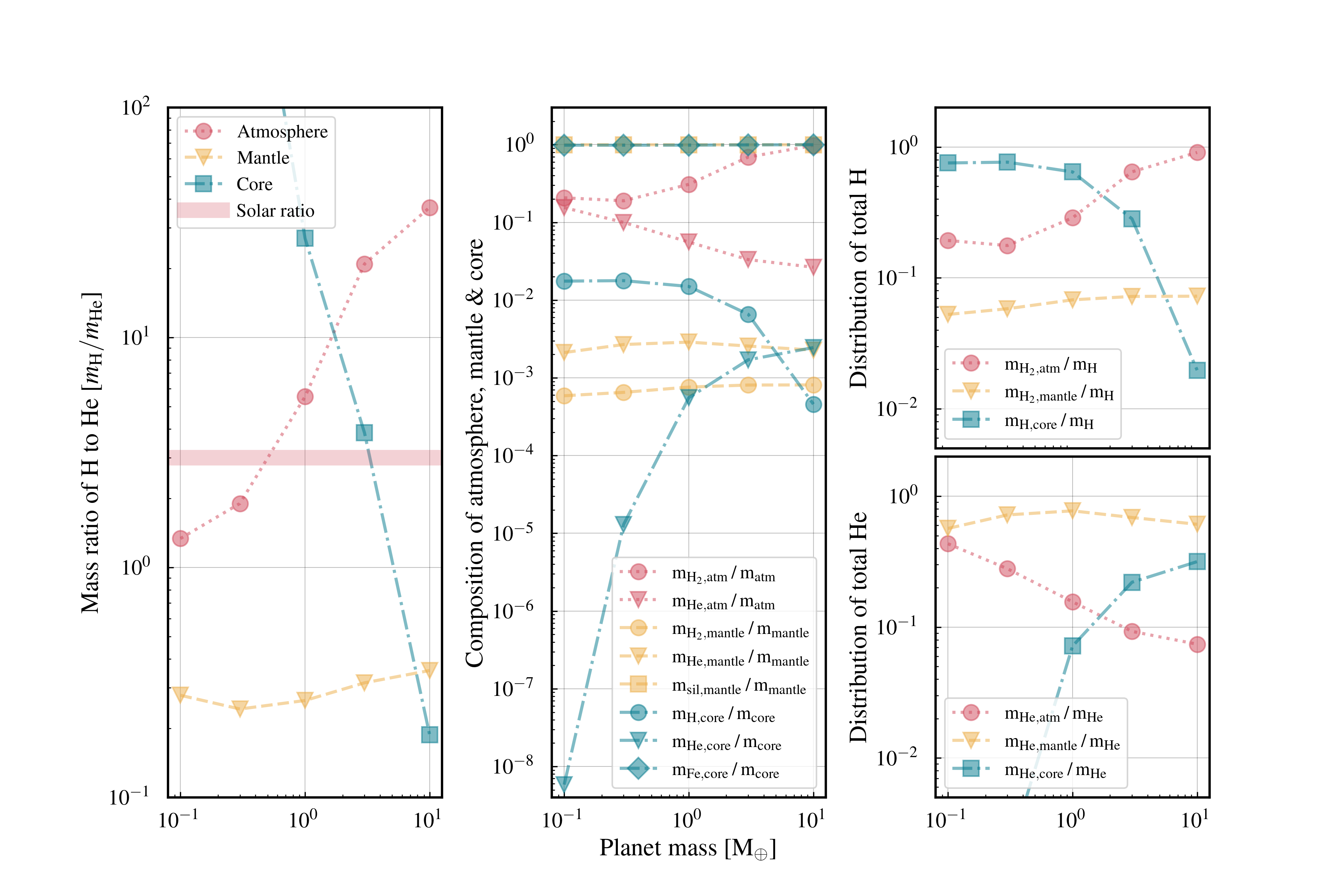}
    \caption{Composite figure illustrating the contrasting equilibrium partitioning of H and He as a function of planet mass. The right panels display the distribution of the total H (top) and He (bottom) budgets among atmosphere, mantle, and core. The central panel depicts the compositional makeup of these distinct phases. The left panel shows the H-to-He mass ratios across these phases, with the pink line indicating the Solar reference value. The results indicate that hydrogen is sequestered in planetary cores at low masses but shifts to the atmosphere for Neptune-like planets. In contrast, helium resides primarily in mantles, with the core becoming an increasingly important reservoir at higher masses.
    } 
     \label{fig:trends_w_planet_mass}
\end{figure}

\begin{figure}
\centering
    \includegraphics[width=\textwidth,trim=0 -50 0 0,clip]{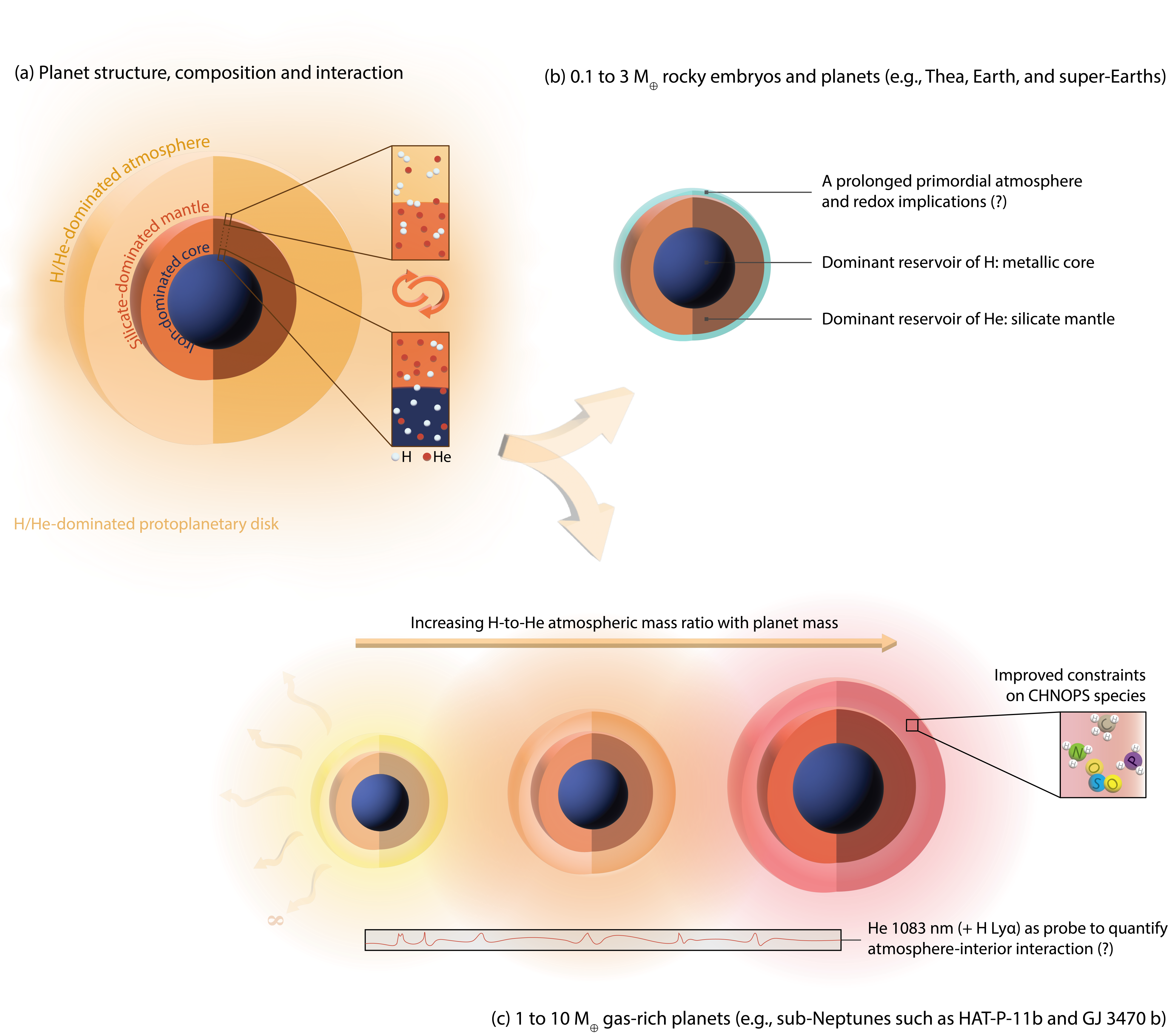}

        \caption{Illustration of atmosphere–mantle–core coupling across rocky embryos to gas-rich planets.
Mantle convection acts as a planetary-scale “conveyor belt,” facilitating chemical communication between the atmosphere and the core. The efficiency of this volatile cycling---potentially probed by the He 1083 nm feature, especially when combined with H Lyman-$\alpha$---depends on the partition coefficients derived in this study. It governs the redistribution of H and He, the timescale of primordial atmosphere retention, aspects of redox balance in secondary atmospheres, and model predictions for the abundances of atmospheric CHNOPS species, ultimately shaping the H-to-He ratio in the atmospheres of Earth- to Neptune-like planets.
    } 
     \label{fig:conclusion_schematic}
\end{figure}

These findings on H and He partitioning between the mantles and cores of planets with Earth-like interiors have broad consequences for the formation, evolution, and composition of planet atmospheres and interiors. These consequences become apparent when we incorporate our partitioning results into a coupled chemical equilibrium–interior model that links atmosphere, mantle, and core~\citep{schlichting2022a,young2023a}; see \textit{Methods}. Our objective is not to construct a complex parameterized framework, but rather to build a reasonable model that captures the potentially far-reaching physical and chemical implications of volatile dissolution in the mantle and core, and highlights the critical need for accurate data quantifying these processes; see~Figure~\ref{fig:trends_w_planet_mass} and~\ref{fig:conclusion_schematic}.

The typical planet forms in the presence of an H and He-rich protoplanetary disk. This view is supported by exoplanet demographics, the composition of Solar System gas-rich planets, isotopic anomalies in Earth and meteorites, and evidence of gas accretion in disks \citep{fulton2017a,helled2020a,mukhopadhyay2019a,vanCapelleveen2025a}. Under such conditions, planets can accrete optically thick H/He-dominated atmospheres and retain them for millions to billions of years, dictating their long-term evolution. For proto-super-Earths and sub-Neptunes, the total H/He mass fraction is typically inferred to be $\sim 0.01 - 10$\% of the planet’s mass~\citep{owen2017a,gupta2019a,lee2015a}, with the exact value depending on the efficiency of cooling during accretion, among other factors~\citep{lee2015a,ginzburg2016a}. For our nominal model, we therefore consider atmospheres composed solely of H and He and that these volatiles in total contribute to 1\% of the planet's mass.

The high temperatures beneath such optically thick H/He atmospheres could sustain fully convective mantles~\citep{ginzburg2016a}. We propose that such a mantle acts as a planetary-scale ``conveyor belt,'' facilitating chemical communication between the atmosphere and the core. A parcel of this mantle near the mantle-atmosphere boundary first equilibrates with the atmosphere, its composition dictated by atmosphere-silicate partition coefficients. This parcel subsequently convects into the deep interior, where it interacts with the core and establishes a new equilibrium, transferring a portion of its atmospheric volatiles. On the one hand, the convective turnover in the young and molten mantles of sub-Earths occurs on timescales of 10$^4$ years~\citep{young2023a}, whereas even for Earth's present-day outer liquid core, the convective timescales are around 10$^3$ years~\citep{buffett2000a}. In either case, convective timescales are orders of magnitude shorter than the 10$^7$ to 10$^{10}$ year lifetimes of primordial atmospheres. A persistent convective cycle would thus efficiently drive the atmosphere, mantle, and core toward a global chemical equilibrium. 

The mantle is thus assumed to be a convecting magma ocean primarily constituted of molten MgSiO$_3$, whereas liquid iron is considered representative of the core. The mantle-to-core mass ratio of 2:1 closely follows Earth. Equilibrium abundances of H and He in these reservoirs are then determined by minimizing the Gibbs free energy of the relevant reactions at pressure–temperature conditions appropriate to the mantle–atmosphere and core–mantle boundaries. The temperature at the mantle-atmosphere boundary is assumed as 2000 K to limit extrapolation of available data on H and He solubility in silicates. Other pressure and temperature conditions are determined by solving planetary structure equations as discussed in \textit{Methods}.

% \vspace{2mm}
\subsection*{{Rocky planetesimals to Earth-like planets: hydrogen-rich cores, helium-rich mantles, \\and reducing atmospheres}}
Immersed in an H-He dominated protoplanetary disk, planetary embryos or planetesimals with $\sim$0.1-0.5 Earth masses, develop H-rich cores and He-dominated mantles and atmospheres (Figure~\ref{fig:trends_w_planet_mass}). Evidence for such capture of disk gas into Earth's building blocks has been widely discussed~\citep{mukhopadhyay2012a, mukhopadhyay2019a}. For the conditions considered here, nearly 70-80\% of the accreted H resides in the cores of such embryos. As suggested for Earth, a large amount of this dissolved H and He may get locked in the cores and mantles of accreted terrestrial planets~\citep{tagawa2021a,young2023a,okuchi1997a,suer2023a}, while lighter atmospheric volatiles would be lost due to late-stage giant impacts~\citep{biersteker2021a}.

Whether these small embryos indeed mark the final stage of equilibration for present-day terrestrial planets is uncertain. Given that the accreted H/He atmosphere could allow a planetary-scale ``conveyor belt'' via a hot, convecting mantle, the last stage of equilibration could occur when today's rocky planets were nearly Earth-mass or even larger but before they lost their primordial atmospheres. Nevertheless, even for 0.5-3 Earth mass rocky planets, most of the accreted H still resides in the core, with atmosphere retaining an increasing fraction with larger planet masses (Figure~\ref{fig:trends_w_planet_mass}). The mantle, however, emerges as the sole, dominant reservoir of He by nearly an order of magnitude. 

In both evolutionary scenarios, as the planet cools and the primary atmosphere is being stripped away by various escape processes over millions to billions of years~\citep{owen2017a,gupta2020a,biersteker2021a,zahnle1986a,hunten1987a}, the dissolved volatiles can exsolve from the crystallizing interior~\citep{yuan2023a}. This outgassing would substantially replenish the primary atmospheres, extending the period during which the planet is shrouded in H and He and challenging the inferences about the efficiency of atmospheric loss processes~\citep{chachan2018a}.

Alternatively, depending on the extent of exsolution, planetary cooling, and atmospheric escape, the dissolved H can instead contribute to reducing surface conditions in planets with secondary CO$_2$- or N$_2$-dominated atmospheres \citep{gu2024a,deng2020a,krissansentotton2024a,kasting2003a,tian2005b} or thin, long-lived envelopes of H/He \citep{misener2021a}. Such reducing conditions are critical for prebiotic chemistry in rocky worlds \citep{urey1952a,miller1953a,kitadai2018a} and could even facilitate greenhouse warming, contributing to habitable conditions on those with thin, long-lived H/He atmospheres \citep{wordsworth2013a,pierrehumbert2011a,seager2020a,mollous2022a}. A shifting balance between atmospheric escape and outgassing from this deep H reservoir as the interior cools and solidifies may thus help oxidize the interior and reduce the atmosphere, influencing surface redox evolution and the potential for prebiotic chemistry---a process that may be common in the early lives of rocky 0.5–3 Earth-mass planets as a natural by-product of planet formation.

% \vspace{2mm}
\subsection*{Sub-Neptunes: dependence of atmospheric H$_2$ and He abundance on planet mass}

Sub-Neptunes ($\sim$1 to 20 Earth masses) are the planets that retain their primordial H/He atmospheres~\citep{lopez2013a,owen2013a,owen2017a,ginzburg2018a,gupta2019a}. We find that the H and He abundance in the atmosphere and core for such planets is a strong function of planet mass (Figure~\ref{fig:trends_w_planet_mass}. As planet mass increases from 1 to 10 M$\oplus$, H shifts from being predominantly stored in the core to being concentrated in the atmosphere. Conversely, He becomes increasingly siderophilic at higher pressures, causing more of it to be sequestered in the core with increasing pressure at the core-mantle boundary, or equivalently, with increasing planet mass. This divergent partitioning causes the atmospheric H/He ratio to change by more than an order of magnitude across the sub-Neptune mass range. 

This finding directly impacts inferences of planetary composition from mass-radius measurements. Previous studies, based on extrapolations from low-pressure experiments, suggested H dissolution could significantly inflate planetary cores~\citep{schlichting2022a}. Our results corroborate this for Earth-mass planets, but they imply that such a core-density deficit driven by H is unlikely for planets more than a few times the mass of Earth.

Our work also provides a new insights into the emergence of He-dominated planets. Such planets are predicted near the upper edge of the radius valley~\citep{hu2015a,malsky2023a} and in the habitable zones of Sun-like stars~\citep{lammer2025a}. However, prior models do not account for the preferential dissolution He in the deep interior. Our findings suggest that because most He is initially sequestered in planetary interiors, subsequent H loss and He exsolution could make H-loss more efficient and atmospheres more He-rich than predicted at later evolutionary stages. This scenario could explain the puzzling non-detections of Ly$\alpha$ around certain sub-Neptunes that are otherwise expected to be losing their atmospheres at a high rate~\citep{dossantos2023a}.

The scenario is more complex for lower-mass planets: $\sim$1-5 Earth masses. For these bodies, a larger reservoir of dissolved H could replenish the atmosphere over long timescales, likely preventing the emergence of He-dominated atmospheres. The ultimate atmospheric composition will be further complicated by mixing with heavier volatiles such as water and ammonia~\citep{benneke2024a,schlichting2022a,gupta2025a,lichtenberg2025a}, highlighting a rich area for future study.

% \vspace{2mm}
\subsection*{Relative abundance and observability of CHNOPS-bearing species in the \\H$_2$/He-dominated atmospheres}

Sub-Neptune atmospheres are among the easiest to characterize because of their large scale heights~\citep{benneke2024a,lichtenberg2025a}. Observational studies often seek constraints on species containing life-essential elements (CHNOPS), such as CH$_4$, CO$_2$ and NH$_3$. Current constraints, when guided by physically motivated models, suggest abundances typically below 10\% by mass, with H$_2$/He forming the dominant background. Our results refine the theoretical basis for such estimates by providing tighter limits on the H$_2$/He baseline, thereby enabling more accurate assessments of CHNOPS abundances and more targeted searches in future observational surveys.

To illustrate, we evaluate the exchange coefficient for H partitioning between mantle and core, K$_{D}^\mathrm{H_{2, mantle} \;\rightleftharpoons\; H_{core}}$. Our calculations yield a K$_{D}$ value more than 100 times larger than those derived from the low-pressure experimental data ($<$10 GPa) used in prior studies~\citep{schlichting2022a,young2023a,nixon2025a}.  This discrepancy arises because hydrogen becomes significantly more siderophilic at the immense pressures relevant to sub-Neptune interiors. This implies that the atmospheric H$_2$ inventory could be overestimated by more than two orders of magnitude in earlier models, leading to correspondingly lower predicted abundances of CHNOPS species in sub-Neptune atmospheres---unless these planets accreted proportionally more H.

% \vspace{2mm}
\subsection*{Helium 1083 nm absorption line as an observational probe of atmosphere-interior interaction}

A key assumption of our planetary chemical equilibrium model is that the entire liquid core equilibrates with the mantle and atmosphere. The degree of such coupling is poorly constrained, yet this caveat offers a unique opportunity for exoplanet science.

As outlined above, mantle convection may act as a ``conveyor belt" for volatiles, continuously cycling volatiles between atmosphere, mantle, and core. The efficiency of this conveyor mechanism, however, is unknown. On Earth and other Solar System bodies, noble gases are widely used as inert tracers of volatile cycling and differentiation~\citep{burnard2013a,marty2012a,porcelli2002a}. He is unique among these in being readily detectable in exoplanets through its 1083 nm absorption line~\citep{dossantos2023a,seager2000a,oklopcic2018a}. We therefore propose that population-level statistics of the He 1083 nm feature, particularly when analyzed alongside the H Lyman-$\alpha$ line, can be used as a probe to quantify the degree of core-mantle-atmosphere interaction across the sub-Neptune population. Ongoing and future surveys---such as TUNES~\citep{vissapragada2024a}, the WINERED Helium Consortium, and STEL$\alpha$~\citep{loyd2024a}---could shed light on this.

While atmospheric fractionation is relatively well understood~\citep{tian2005b,hu2015a,malsky2023a,lammer2025a,zahnle1990a}, the efficiency of atmosphere-mantle-core interaction is not. As in the various scenarios discussed earlier, He abundance could change significantly depending on the planetary properties. Observationally, if such trends are absent or weaker than predicted, this would directly constrain the efficiency of volatile transport through mantles and the interactivity of cores. This provides an exciting observational window into the core's role in shaping planetary atmospheres, from the composition of secondary atmospheres to the relative abundance of CHNOPS species in H$_2$/He-dominated envelopes.

\vspace{3mm}
\section*{{Conclusion}}

In this work, we present first-principles calculations that advance our understanding of how hydrogen and helium---the universe's most abundant elements---interact with the rock and iron that form the bulk of terrestrial and sub-Neptune planets. Our key finding is that H and He exhibit divergent partitioning behaviors under the extreme conditions of planetary interiors: H becomes strongly siderophilic, sequestering in the core, while He remains preferentially in the silicate mantle.

These results suggest fundamental revisions to current assumptions. Rocky planets like Earth may have been assembled from building blocks with H-rich cores and He-rich mantles, providing a natural mechanism for generating the reducing conditions in early lives of rocky planets thought necessary for the origin of life. For sub-Neptunes, the atmospheric H-to-He mass ratio emerges as a strong function of planet mass. Smaller sub-Neptunes develop H-enriched cores, whereas the more massive ones do not. Conversely, although He largely remains in the mantle, its solubility in the core increases with planet mass. Consequently, the lower-mass sub-Neptunes possess substantially smaller H$_2$/He atmospheric reservoirs than previously predicted. Unless their total accreted H/He budgets are orders of magnitude larger than current estimates, the abundances of CHNOPS-bearing molecules such as H$_2$O and CH$_4$ must be correspondingly higher. These deep volatile reservoirs may dictate the emergence of helium-dominated atmospheres near the upper edge of the radius valley and within the habitable zones of Sun-like stars.

Finally, He offers a direct observational test: its 1083 nm absorption line, especially when analyzed alongside the H Lyman-$\alpha$ line, provides a tracer of atmosphere–interior exchange, enabling population-level constraints on the efficiency of the mantle “conveyor belt” that couples atmospheres to cores. Ultimately, understanding the partitioning of these simple elements is not a niche problem---it is foundational to connecting the observed diversity of exoplanets to their formation, evolution, and potential for hosting life, underscoring the need to integrate first-principles physics into our models and interpretations of worlds beyond our own.

\newpage
% \newpage

\section*{{APPENDIX}}

\vspace{5mm}
\section*{{Methodology}}\label{sec:methodology}

We calculated partition and exchange coefficients for H and He partitioning between silicate melts and liquid iron by calculating their free energies, and consequently, the chemical potentials of their mixtures at a series of concentrations of H and He~\citep{luo2024a,li2020a,li2022a}. To this end, we performed \textit{ab-initio} molecular dynamics (AIMD) coupled with thermodynamic integration (TI).

\vspace{2mm}
\subsection*{Molecular dynamics simulations}

We performed AIMD simulations on the H-He-Fe-MgSiO$_3$ system using the Vienna \textit{ab initio} Simulation Package (VASP). The projector-augmented wave (PAW) method was used~\citep{kresse1993a,kresse1996a,kresse1996b,kresse1999a}. The valence configurations were Fe (3$p^6$ 3$d^7$ 4$s^1$), Mg (2$p^6$ 3$s^2$), Si (3$s^2$ 3$p^2$), O (2$s^2$ 2$p^4$), He (1$s^2$), and H (1$s^1$). For the exchange-correlation potential, we used the PBEsol approximation~\citep{perdew2008a}, which has been shown to reproduce experimental data reliably~\citep{scipioni2017a,holmstrom2018a}. Thermal equilibrium between ions and electrons was assumed via the Mermin functional~\citep{mermin1965a,wentzcovitch1992a}.  

For the thermodynamic integration, the supercells contained 64 Fe atoms for iron liquids and 160 atoms, or 32 formula units, of MgSiO$_3$ liquids. For H- and He-bearing compositions, H and He atoms were inserted into these supercells. Each trajectory was propagated for 5,000-40,000 steps with a timestep of 0.25 fs while ensuring equilibration. We sampled the Brillouin zone at the $\Gamma$ point and used plane-wave energy cutoffs of 600 eV for iron and 800 eV for silicate melts. 
We applied a pressure and energy correction using a $2 \times 2 \times 2$ Monkhorst--Pack $k$-point mesh to account for the relatively limited supercell sizes.

We performed canonical ensemble (\textit{NVT}) and isothermal--isobaric (\textit{NPT}) simulations with temperature controlled by a Nos\'e-Hoover~\citep{nose1984a,nose1991a,hoover1985a,frenkel2001a} and Langevin thermostat~\citep{allen2017a,hoover1982a,evans1983a}, respectively, and for the latter, pressure controlled by the Parinello-Rahman barostat~\citep{parrinello1980a,parrinello1981a}. To determine equilibrium volumes, we carried out \textit{NPT} simulations at the target pressures followed by \textit{NVT} simulations at multiple volumes around the target pressure and fitted the resulting pressure-volume data to obtain an equation of state. Radial distribution functions and diffusion coefficients were computed to verify that the simulated systems remained in the liquid state. Statistical errors of the internal energies or any other simulation-derived quantity were estimated by block averaging~\citep{flyvbjerg1989a}, and for uncertainty propagation, we used a Monte Carlo scheme.

\vspace{2mm}
\subsection*{Thermodynamic integration}

We determined the Gibbs free energies of the metallic and silicate melts by TI from the ideal gas reference. To assess the robustness of this choice, we also employed the Weeks–Chandler–Andersen (WCA) fluid as the reference for the liquid phase and found the resulting free-energy differences to be negligible. Specifically, we integrated the difference in internal energies between the melt ($U_1$) and the reference state ($U_0$) with respect to the coupling parameter $\lambda$:  
\begin{equation}
\Delta F_{0 \to 1} = \int_0^1 d\lambda \,\langle U_1(R) - U_0(R)\rangle_\lambda.
\end{equation}

To mitigate numerical instabilities caused by close contacts between non-interacting atoms at $\lambda = 0$, we employed a variable transformation \citep{dorner2018a}, with 
\begin{equation}
\lambda(x) = \left(\frac{x+1}{2}\right)^{1/(1-k)},
\end{equation}
which recast the integral as  
\begin{equation}
\Delta F = \frac{1}{2(1-k)} \int_{-1}^1 f(\lambda(x)) \,\lambda(x)^k \, dx,
\end{equation}
where we adopted $k = 0.8$. The integral was evaluated using an 8-point Gauss-Lobatto quadrature.

The Gibbs free energy of the ideal gas was computed as  
\begin{equation}
G^{\mathrm{ig}} = F^{\mathrm{ig}} + PV 
= -k_\mathrm{B}T \sum_i \ln\!\left(\frac{V^{N_i}}{N_i! \,\Lambda_i^{3N_i}}\right) + PV,
\end{equation}
where $k_\mathrm{B}$ is Boltzmann’s constant, $T$ is the temperature, $P$ the pressure, $V$ the volume, $N_i$ the number of atoms of element $i$,  and $\Lambda_i = h/(2\pi M_i k_\mathrm{B}T)^{1/2}$ is the thermal wavelength of species $i$, with $h$ as the Planck’s constant and $M_i$ the atomic mass.  

All calculations were calibrated to the target pressure using the thermodynamic relation  
\begin{equation}
G(P_2, T) - G(P_1, T) = \int_{P_1}^{P_2} V \, dP.
\end{equation}

To evaluate free energy differences between two temperatures along a fixed pressure, we estimated and integrated the ensemble-averaged enthalpy. The instantaneous enthalpy is defined as $H = \mathcal{U} + PV$, and its average was sampled within the isothermal-isobaric ensemble. Integrating the enthalpy contribution between a reference temperature $T_0$ and a target temperature $T_1$ yielded the Gibbs free energy difference:  
\begin{equation}
\frac{G(P,T_1)}{k_\mathrm{B}T_1} - \frac{G(P,T_0)}{k_\mathrm{B}T_0} 
= \int_{T_0}^{T_1} \frac{1}{k_\mathrm{B}T^2}\,\langle H \rangle_{P,T}\, dT.
\end{equation}

The enthalpy averages were obtained from \textit{ab initio} molecular dynamics simulations. $NPT$ ensemble simulations were employed to estimate accurate volumes at each discrete integration point. The integrand was then estimated using $NVT$ ensemble simulations spanning 20,000 to 40,000 timesteps. The first 5,000-10,000 steps were discarded for reported values.

\vspace{2mm}
\subsection*{Estimating the chemical potentials, partition coefficients, and equilibrium constants}

At thermodynamic equilibrium between the coexisting metallic and silicate liquids---liquid Fe and molten MgSiO$_3$, respectively---the chemical potentials of each volatile species $i \in {\mathrm{H},\mathrm{He}}$ must be equal in both phases,
\begin{equation}
\mu_i^{\mathrm{Fe}}(P,T,x^\mathrm{Fe}_{i})=\mu_i^{\mathrm{MgSiO_3}}(P,T,x^\mathrm{MgSiO_3}_{i}),
\label{eq:mu_eq}
\end{equation}
where $\mu_i^{j}$ is the chemical potential of species $i$ in phase $j \in {\mathrm{Fe},\mathrm{MgSiO_3}}$, expressed as a function of pressure ($P$), temperature ($T$), and concentration of species $i$ in the respective phases, $x^{j}_{i}$. The chemical potential can be written as the partial molar Gibbs free energy,
\begin{equation}
\mu^{j}_{i}(P,T,x_i^j) = \bar{\mu}^{j}_i(P,T,x_i^j) - T S_{\mathrm{mix}}^{j} =  \left( \frac{\partial G^j_{i}}{\partial x} \right)_{(P,T,x_i^j)}.
\label{eq:mu_defn}
\end{equation}

Here $T S_{\mathrm{mix}}^{j}$ is the ideal gas mixing entropy contribution whereas $\bar{\mu}^{j}_{i}$ is the non-ideal (excess) part. Physically, $\mu_i^j$ corresponds to the change in molar Gibbs free energy $G^j$ when an infinitesimal amount of species $i$ is introduced into phase $j$ at fixed $P$ and $T$. These quantities were determined using thermodynamic integration combined with ab initio molecular dynamics (TI-AIMD), as described above. For simplicity, $x_i^j$ is hereafter denoted as $x$.

The Gibbs free energy of a system at any $P$, $T$, and $x$ can be defined as
\begin{equation}
    G^{j}_{i}(P,T,x) = \bar{G}^{j}_i(P,T,x) - T S_{\mathrm{mix}}^{\mathrm{system}},
    \label{eq:G_defn}
\end{equation}
where $\bar{G}$ is the excess free energy and $S_{\mathrm{mix}}^{\mathrm{system}}$ is the ideal entropy of mixing for our four systems $\in$ $j_{(1-x)} i_x$. From~Equations~(\ref{eq:mu_defn}) and~\ref{eq:G_defn}, and the Gibbs-Duhem relation~\citep{li2022a},
\begin{align}
\bar{\mu}_{i}^{j} (P,T,x) &= 
\bar{G}^{j}_{i} (P,T,x) + (1-x)\,\frac{\partial \bar{G}^{j}_{i} (P,T,x)}{\partial x} \quad \text{and}\\
S_{\mathrm{mix}}^{i\mathrm{\,in\,}j} &= 
S_{\mathrm{mix}}^{\mathrm{system}} + (1-x)\,\frac{\partial S_{\mathrm{mix}}^{\mathrm{system}}}{\partial x}.
\label{eq:S_mix}
\end{align}

We found that $\bar{G}^{j}_{i}$, defined per formula unit with respect to $x$ for our systems $j_{(1-x)} i_x$, varies linearly with $x$:
\begin{equation}
    \bar{G}^{j}_{i}(P,T,x) = \mathrm{a} \; + \; \mathrm{b}\,x
\end{equation}
where a and b are regression parameters obtained from least-squares fitting. Thus, $\bar{\mu}^{j}_{i}$ reduces to (a + b). 

The ideal mixing entropy for the systems $j_{(1-x)} i_x$ was:
\begin{align}
S_{\mathrm{mix}}^{\mathrm{Fe}_{1-x}{i}_x} &=
-k_B \left[\, x \ln x + (1-x)\ln(1-x)\,\right] \quad \text{and}\\
S_{\mathrm{mix}}^{(\mathrm{MgSiO}_3)_{1-x}{i}_x} &=
-k_B \left\{\, x \ln\!\left(\tfrac{x}{5-4x}\right) 
+ (1-x) \left[\, 2\ln\!\left(\tfrac{1-x}{5-4x}\right) 
+ 3\ln\!\left(\tfrac{3-3x}{5-4x}\right) \right] \right\}.
\end{align}

Therefore, using~Equation~(\ref{eq:S_mix}), $S_{\mathrm{mix}}^\mathrm{i\,in\,j}$ reduced to
\begin{align}
S_{\mathrm{mix}}^{i\mathrm{\,in\,Fe}} &=
-k_B\,\mathrm{ln}(x) 
\quad \text{and}
\\
S_{\mathrm{mix}}^{i\mathrm{\,in\,MgSiO_3}} &=
-k_B\,\mathrm{ln}\left(\frac{x}{5-4x}\right).
\end{align}

These equations were then substituted back in~Equation~(\ref{eq:mu_defn}) to estimate $x^j_i$, i.e., the equilibrium mole fractions of H and He in Fe, i.e., the metallic phase (`met') and MgSiO$_3$, i.e., the silicate phase (`sil'). The mole-fraction-based exchange coefficient $K_{\text{D, }i}^{\text{met/sil}}$ and the weight-based partition coefficient D$_i^{\text{met/sil}}$ for the partition of H and He between metallic and silicate phases are then 
\begin{align}
    K_{\text{D, }i}^{\text{met/sil}} &= \frac{x_{i}^\text{Fe}}{x_{i}^{\text{MgSiO}_3}} = \frac{x_{i}^\text{met}}{x_{i}^{\text{sil}}}
    \quad \text{and}\label{eq:K_D_defn}\\
D_i^{\text{met/sil}} &= \frac{w_{i}^\text{Fe}}{w_{i}^{\text{MgSiO}_3}}  = \frac{w_{i}^\text{met}}{w_{i}^{\text{sil}}},
\end{align}
where $w$ indicates the corresponding equilibrium weight fractions. To estimate the 1-$\sigma$ uncertainties for partition coefficients and other variables, we used a Monte Carlo uncertainty-propagation scheme that assumes an asymmetric Gaussian distribution of inputs and uses rejection sampling to enforce physically motivated bounds.

\vspace{2mm}
\subsection*{Two-phase simulations}
In complement to the free-energy based TI-AIMD approach, we also perform two-phase \textit{ab initio} molecular dynamics simulations (2P-AIMD) to gain additional insights into the lithophilic or siderophilic nature of H and He~\citep{gupta2025a,xiao2018a}. These simulations are initiated as domains of equilibrated molten silicate (MgSiO$_3$-dominated) and metallic (Fe-dominated) phases, with H or He atoms mixed-in and joined at a planar interface. These simulations comprise 500-700 atoms in total: 128 atoms of Fe, and 64 formula units ($\equiv$ 320 atoms) of MgSiO$_3$, and either 240 H or 80 He atoms. The system is then equilibrated at a target temperature and pressure: 6500 K and 250 GPa by evolving it for 10 ps with a 0.5 fs time-step such that the silicate and metallic phases reach a dynamic equilibrium, i.e., when the chemical potentials of the two phases equalize. 

For each trajectory, the instantaneous coarse-grained atomic density $\rho(\mathbf{r})$ was computed using a Gaussian smoothing length of 0.25 nm~\citep{willard2010a,sega2018a} and projected along the axis normal to the interface to obtain $\rho(z)$ for the simulation cell. The resulting profiles were fitted with a theoretically-motivated hyperbolic-tangent function to identify the Gibbs dividing surfaces separating the silicate and metallic phases~\citep{cahn1958a,widom1982a}. H and He counts within each region, averaged over the last 2 ps, yielded equilibrium concentrations and corresponding metal–silicate partition coefficients; see Figure 9. The partition coefficients derived from this approach are, however, influenced by finite-size effects arising from the system dimensions and the limited number of atoms.

\vspace{2mm}
\subsection*{Modeling atmosphere-mantle-core interaction}
To demonstrate the potential impact of the partition coefficients on a planet's formation, evolution, and atmospheric composition, we developed a self-consistent model for interaction across atmosphere-mantle-core~\citep{schlichting2022a,young2023a,bower2025a}.

\vspace{1mm}
\subsubsection*{Planet atmosphere-interior structure}
Following previous studies, our fiducial planet model has three layers: (1) an optically thick atmosphere, (2) a mantle, and (3) a core. We assume that the atmosphere is dominated by H$_2$ and He alone. The mantle is a convecting magma ocean that is nearly adiabatic and is MgSiO$_3$-dominated, whereas the core is primarily composed of liquid iron.

For the temperature and pressure conditions at the mantle-atmosphere boundary, i.e., MAB, we assume a fixed T$_\mathrm{MAB}$ = 2000 K and
\begin{equation}
P_{\mathrm{MAB}}
\simeq 10^{2} \, \frac{M_{\mathrm{atm}}}{M_{p}}
\left( \frac{M_{p}}{M_{\oplus}} \right)^{2/3} \;\mathrm{GPa},
\label{eq:P_atm}
\end{equation}
where $M_p$ and $M_\mathrm{atm}$ are the planet mass and atmospheric mass, respectively~\citep{schlichting2022a}. While the trends noted in the main text are not sensitive to the exact value of T$_\mathrm{MAB}$, we choose a value of 2000 K to avoid extrapolation of the laboratory and simulation where possible. H and He solubility in silicates is only available for temperatures around 1500-2500 K~\citep{bouhifd2013a,hirschmann2012a,wang2023a}.

To estimate the internal structure of the planet, we developed a Julia-Python-based code, \texttt{PAI-Solver} (Planet-Atmosphere-Interior-Solver), to solve the partial differential equations of hydrostatic equilibrium in mass coordinates~\citep{seager2007a}:
\begin{align}
\frac{\partial r(m)}{\partial m} &= \frac{1}{4 \pi r^2 \rho}, \\[6pt]
\frac{\partial P(m)}{\partial m} &= - \frac{G m}{4 \pi r^4}, \quad \text{and} \\[6pt]
\rho(m) &= f(P(m),T(m)).
\end{align}

Here $r$ is the radius, $P$ the pressure, and $m$ the enclosed mass. $f$ is the equation of state (EOS) for mantle (MgSiO$_3$) or core (Fe). A phase boundary is enforced such that $\rho = \rho_{\rm mantle}(P,T)$ for $m > m_{\rm core}$ and 
$\rho = \rho_{\rm core}(P,T)$ for $m \leq m_{\rm core}$.

The outer boundary is specified by $(R_{\mathrm{MAB}}, P_{\mathrm{MAB}}, T_{\mathrm{MAB}})$, with $R_{\mathrm{MAB}}$ determined iteratively to satisfy the central boundary condition. The CMB temperature was scaled from the MAB assuming a Grüneisen parameter $\sim$1~\citep{luo2025a}. The planet structure part of \texttt{PAI-Solver} output $R_{\mathrm{MAB}}$, $P_{\mathrm{MAB}}$, $T_{\mathrm{MAB}}$, $\rho_{\mathrm{MAB}}$, the CMB location and state variables, central pressure and density.

For the equation of state for MgSiO$_3$ and Fe, we used the polytrope fits from Seager et al.~\citep{seager2007a}, based on data from Ahrens, Karki et al. and Anderson et al.~\citep{ahrens1995a,karki2000a,anderson2001a}. While these approximations yield errors in radius and pressures of order 10\%, they are suitable for demonstrating the underlying physical and chemical implications of atmosphere-interior interactions, which are typically far more uncertain and unknown, and are the primary concern of this study.

\vspace{1mm}
\subsubsection*{Global chemical equilibrium optimization}
To demonstrate the potential impact of D$_i^{\text{met/sil}}$ on a planet's formation, evolution, and atmospheric composition, we coupled the planet structure solver with a model for chemical interaction across the atmosphere-mantle-core regimes~\citep{schlichting2022a,young2023a}. In particular, we focus on four independent reactions:
\begin{align}
\text{[R1]} \quad & \mathrm{H_{2,atm}}     \;\rightleftharpoons\; \mathrm{H_{2,mantle}}, \\
\text{[R2]} \quad & \mathrm{He_{atm}}      \;\rightleftharpoons\; \mathrm{He_{mantle}}, \\
\text{[R3]} \quad & \mathrm{H_{2,mantle}}  \;\rightleftharpoons\; \mathrm{H_{core}}, \quad \text{and} \\
\text{[R4]} \quad & \mathrm{He_{mantle}}   \;\rightleftharpoons\; \mathrm{He_{core}},
\end{align}
where R1 and R2 occur at the MAB, whereas the latter two at the CMB. For these four reactions R$_q$, where $q\in\{1,2,3,4\}$, the equilibrium conditions can be written as
\begin{equation}
\left[ 
\frac{\Delta G^{\circ}_{\mathrm{R}_q} (P,T,x_{k})}{R_\text{gas}\,T} 
+ \sum_{k \in \mathrm{atm}} \nu_k \ln \!\left(\frac{P}{P^{\circ}}\right) 
\right] 
+ \sum_{k} \nu_k \ln x_k 
= 0 \;\;\text{or}
\label{eq:eq_cond1}
\end{equation}

\begin{equation}
\left[ 
- \ln \!\left( K_{\mathrm{eq,\,R_q}} (P,T,x_{k}) \right) 
+ \sum_{k \in \mathrm{atm}} \nu_k \ln \!\left(\frac{P}{P^{\circ}}\right) 
\right] 
+ \sum_{k} \nu_k \ln x_k 
= 0,
\label{eq:eq_cond2}
\end{equation}
where $\Delta G^{\circ}_{\mathrm{R}_q}$ and $K_{\rm eq,\,R_q}$ are the standard-state Gibbs free energy change and the equilibrium constant, respectively, and are functions of the pressure ($P$) and temperature ($T$) at the relevant interface (MAB or CMB) and the mole-fractions {$x$} of all participating species. $R_\text{gas}$ is the gas constant and $P^\circ$ is the pressure at the standard state, chosen as 1 bar. The first summation is over \textit{k}, which denotes the different species $k$ in their respective host phases (atmosphere, mantle, or core), and $\nu_k$ are their respective stoichiometric coefficients in the reaction $q$. The latter sum over ``atm" accounts for the atmospheric or gaseous species. The compositional and pressure contributions for the gaseous species are thus separated, replacing partial pressure (ideal fugacity) by the product $(x\,P)$ for atmospheric species.

This formulation links the standard-state free energies or equilibrium constants with the $({P,T,x})$ conditions in the different reservoirs, providing the thermodynamic framework for H–He distribution implied by reactions [R1-R4].

In addition, we impose conservation constraints to ensure that the total abundance of each element is preserved across the three reservoirs or phases $j \in \{\text{atmosphere, mantle, core}\}$. For each element $i \in \{\text{H,\,He}\}$, a constraint on its mass or, equivalently, moles, can be written as
\begin{equation}
n_i - \sum_{j} \sum_{k} \eta_{i,k}^j \, x_{k}^j \, N_j = 0 ,
\label{eq:mole_cons}
\end{equation}
where $n_i$ is the total moles of element $i$ in the planet, $\eta_{i,k}^j$ is the number of atoms of element $i$ contained in component $k$ of phase $j$ (e.g., 2 for H$_2$ component of H in mantle), $x_{k}^j$ is the mole fraction of component $k$ within phase $j$, 
and $N_j$ is the total number of moles of phase $j$.  We assume that MgSiO$_3$ and Fe do not participate in chemical exchange.

Further, the mole fractions within each phase are constrained to sum to unity. Thus, for the three phases, we have
\begin{equation}
1 - \sum_j x^{j}_{k} = 0.
\label{eq:mole_frac_cons}
\end{equation}

Together with~Equation~(\ref{eq:P_atm}) that relates atmospheric pressure with the mass or moles of gases in the atmosphere, the four equations from the equilibration constraint (Equation~\ref{eq:eq_cond1}), the two from the conservation of moles (Equation~\ref{eq:mole_cons}) and three from the summation constraint mole fractions, ensure thermodynamic consistency across the atmosphere, mantle and core. We assume that the P$_\text{CMB}$ and T$_\text{CMB}$ do not change due to the dissolution of volatiles in the core or mantle.

\vspace{3mm}
\subsubsection*{Bayesian inference framework}
We employed a Bayesian inference framework to find the optimal solution to the aforementioned system of equations. Bayes’ theorem allows us to infer the posterior distribution $\mathcal{P}({\bf \Phi} |\mathbf{D},M)$ of the model parameters $\bf \Phi$ for a given model $M$, conditioned on input constraints $\mathbf{D}$: 
\begin{align}
    \mathcal{P}({\bf\Phi}|\textbf{D}, M) &= \frac{\mathcal{P}(\textbf{D}|{\bf\Phi}, M) \times \mathcal{P}({\bf\Phi}|M)}{\mathcal{P}(\textbf{D}|M)}.
\end{align}

Here $\mathcal{P}(\mathbf{D}|{\bf \Phi},M)$ is the likelihood of obtaining the data for a particular choice of parameters and model, $\mathcal{P}({\bf \Phi}|M)$ is the prior distribution encoding any pre-existing knowledge of the parameters, and the evidence is defined as
\begin{align}
    \mathcal{P}(\textbf{D}| M) &= \int_{\forall {\bf\Phi}} {\mathcal{P}(\textbf{D}|{\bf\Phi}, M) \times \mathcal{P}({\bf\Phi}|M)}\;\text{d}{\bf\Phi}.
\end{align}

In the present work,
\begin{enumerate}
\item $M$ is the coupled atmosphere–mantle–core interaction model described above, 
\item $\mathbf{D} = {D_i}$ denotes the 10 model constraints arising from~Equations~(\ref{eq:P_atm}),~(\ref{eq:eq_cond1}),~(\ref{eq:eq_cond2}),~(\ref{eq:mole_cons}) and~(\ref{eq:mole_frac_cons}) and that depend on parameters such as the planetary mass ($M_p$), Grüneisen parameter, equilibrium temperature (T$_{eq}$), mass fraction of volatiles, silicate and metallic melts, primordial H-to-He mass ratio, T$_\text{MAB}$ and importantly, $\Delta G^{\circ}_{\mathrm{R}_q} (P,T,x)$ / $K_{\rm eq,\,R_q} (P,T,x)$ for all reactions at the relevant temperatures and pressures, 
\item ${\bf \Phi} = {\Phi_i}$ includes the unknown model parameters, such as the mole fractions of all the component species in their respective phases, the mass fractions of the respective phases, and the P$_\text{MAB}$.
\end{enumerate}

We assume all ${\Phi_i}$, except P$_\text{MAB}$, are log-uniformly distributed on $\in [0,1]$. For P$_\text{MAB}$, we assume a similar distribution $\in [\text{1 bar, 100 GPa}]$. For each constraint $D_i$, uncertainties are assumed to be Gaussian, with the likelihood centered at zero and standard deviation $\sigma_i$ equal to 10$^{-3}$ for~constraints~(\ref{eq:eq_cond1}) and (\ref{eq:mole_frac_cons}), and 10$^{-6} \, \times \, n_i$ for~constraint~(\ref{eq:mole_cons}). The total likelihood is then the product of the individual likelihoods for all $D_i$. This framework provides a statistically rigorous method to explore the entire relevant parameter space and yields posterior distributions for the partitioning of H and He that are consistent with both computational and experimental constraints on partition coefficients and thermodynamic equilibrium in a planet.

To compute this resulting $P({\bf\Phi}|\textbf{D}, M)$, we leveraged the open-source code \texttt{dynesty}~\citep{speagle2020a,koposov2023a}, which uses a Dynamic Nested Sampling algorithm to estimate posteriors and evidences~\citep{skilling2004a,skilling2006a,higson2019a,feroz2009a}.

\vspace{3mm}
\subsubsection*{Estimating $\Delta G^{\circ}_{\mathrm{R}_q}$ and $K_{\rm eq,\,R_q}$}

For the purpose of finding the optimum parameters, we require a continuous function in $(P,T,x_{R_q})$ space for $\Delta G^{\circ}_{\mathrm{R}_q}$ and $K_{\rm eq,\,R_q}$.  For R1, we use the prescription for $K_{\rm eq,\,R_1}$ from Schlichting and Young and Young et al.~\citep{schlichting2022a,young2023a}, based on the experimental results reported in Hirschmann et al.~\citep{hirschmann2012a} at $\sim$1500 K.  For R2, we assume linear interpolation between $K_D$ values at 1600 and 2500 K from the \textit{ab initio} results of Wang et al.~\citep{wang2023a}, which are consistent with the experiments reported by Bouhifd et al.~\citep{bouhifd2013a}. 

$\Delta G^{\circ}_{\mathrm{R}_q}$ and $K_{\rm eq,\,R_q}$ for R3 and R4 are based on our calculations performed across a range of temperatures, pressures and compositions and consistent with the trends from existing experimental~\citep{tagawa2021a,malavergne2019a,clesi2018a,okuchi1997a,bouhifd2013a,matsuda1993a} and computational studies~\citep{li2020a,li2022a,zhang2012a,xiong2021a,wang2022a}. We find that chemical potentials are not sensitive to composition which implies that the activity coefficients are close to unity. We thus assume that the calculated exchange coefficients are equal to the respective equilibrium constants. For R3, the equilibrium constant can be written as
\begin{align} 
K_{\mathrm{eq},\,R_3}  &= 
\frac{\bigl[x_\mathrm{H}^{\mathrm{met}}\bigr]^2}{\bigl[x_\mathrm{H_2}^{\mathrm{sil}}\bigr]} \\[6pt] &= 
\frac{     1 + \left( \dfrac{1}{w_\mathrm{H}^{\mathrm{sil}}} - 1 \right)          
\dfrac{\mu_\mathrm{H_2}}{\mu_{\mathrm{MgSiO_3}}} }
{     \left[         1 + \left( \dfrac{1}{w_\mathrm{H}^{\mathrm{met}}} - 1 \right)              
\dfrac{\mu_\mathrm{H}}{\mu_{\mathrm{Fe}}}     \right]^2 }, 
\end{align}
whereas for R4, $K_{\text{eq, }R_4} = K_{\text{D,  He}}^{\text{met/sil}}$ as defined in~Equation~(\ref{eq:K_D_defn}).

To obtain compact, differentiable forms for the equilibrium constants  $K_{\text{eq},R_3}$ and $K_{\text{eq},R_4}$, we employ symbolic-regression~\citep{udrescu2020a}, a machine learning approach as implemented in PySR~\citep{cranmer2020a}.  Training data for $\ln K_{\text{eq},R_q}$ are generated from free energy calculations  as functions of $P$, $T$, and $x$. Inputs are non-dimensionalized, the logarithm of $K_{\text{eq}}$ is modeled to enforce positivity, and the search is restricted to smooth mathematical operations. The algorithm explores a large hypothesis space and returns candidate models along the Pareto front, balancing accuracy and complexity. Final coefficients are refined by nonlinear least squares. The resulting closed-form expressions  $f_q (P,T,x)$ serve as continuously differentiable, machine-learned surrogates for $K_{\text{eq},R_q}$ and can be expressed as follows:
\begin{align}
\log \left(K_{\mathrm{eq},\,R_3} \right) &=
    \frac{\log\!\left(x_{\mathrm{H_2}}^{\mathrm{sil}}\right) 
            + \left(T^{0.228} - 0.0081\,P\right)}
            {1.05 + 1.37^{\log \left(x_{\mathrm{H_2}}^{\mathrm{sil}}\right)}}, \quad \text{and}  \\[6pt]
\log \left( K_{\mathrm{eq},\,R_4} \right) &= 
    -\,\frac{0.0678\,T + 44.5}{P}.
\end{align}

\newpage

\section*{{Acknowledgements}}
A.G. acknowledges support from the Heising–Simons Foundation through the 51 Pegasi b Fellowship, and from Princeton University through the Harry H. Hess Fellowship and the Future Faculty in Physical Sciences Fellowship. A.G. also acknowledges support from the National Aeronautics and Space Administration (NASA) under grant HST-GO-17804.011-A. A.G. and A.B. acknowledge support from the Center for Matter at Atomic Pressures (CMAP), a National Science Foundation (NSF) Physics Frontier Center under Award PHY-2020249. J.D. acknowledges support from the NSF under Award EAR-2242946.

The calculations presented in this article were performed on computational resources managed and supported by Princeton Research Computing, a consortium of groups including the Princeton Institute for Computational Science and Engineering (PICSciE) and the Office of Information Technology's High Performance Computing Center and Visualization Laboratory at Princeton University.

\newpage

\bibliographystyle{unsrtnat} % or plainnat/abbrvnat

\bibliography{planet_evo}{}

\bibliographystyle{bibstyle_format_2}

% \printbibliography %Prints bibliography

% % ###################################################
% % ###################################################
% % ###################################################

\newpage
% \newpage

\section*{{Supplementary Information}}

\begin{figure}[h]
\centering
    \includegraphics[width=\textwidth,trim=0 0 0 0,clip]{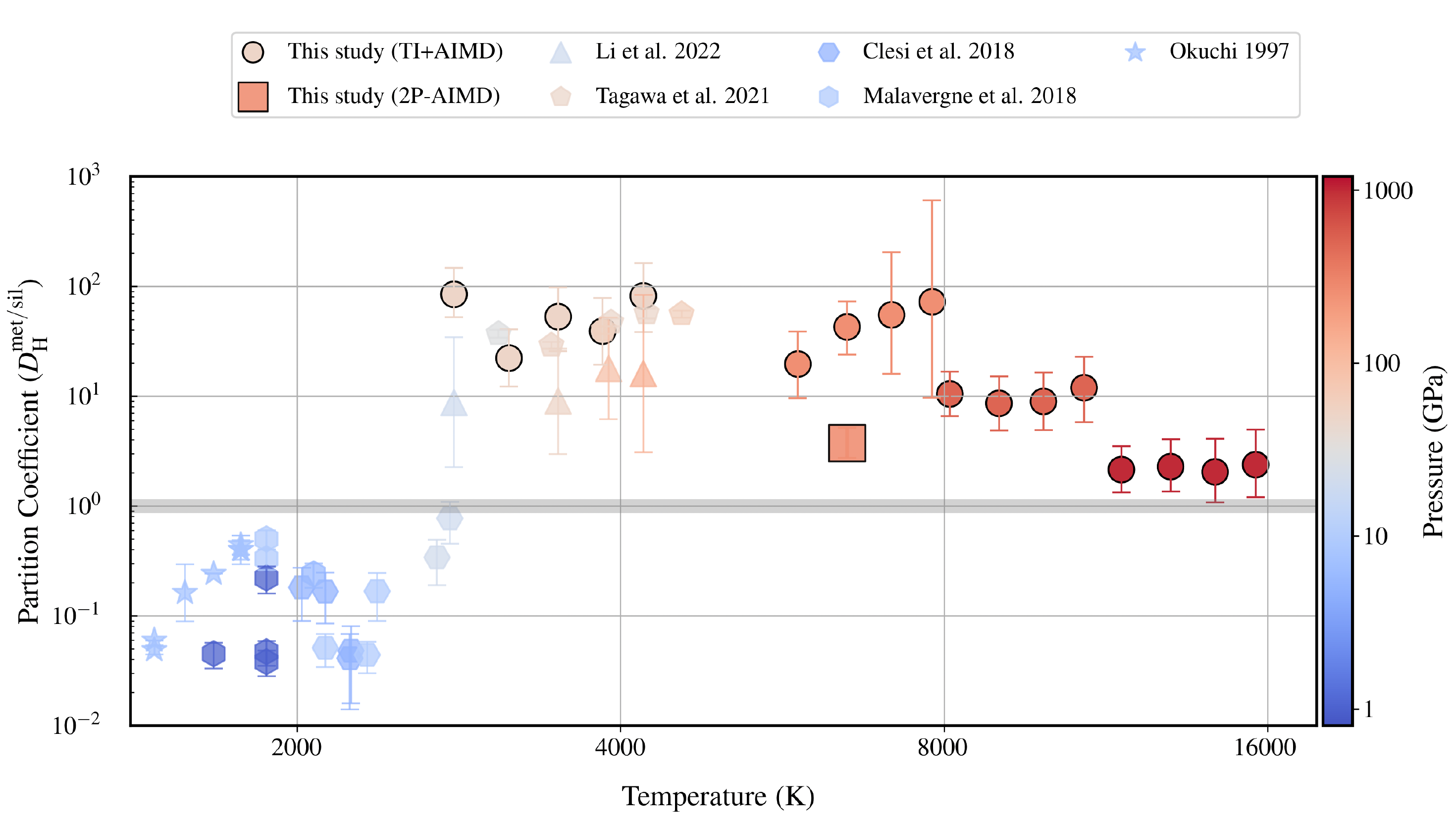}
    \caption{Partition coefficient for hydrogen between silicate and metallic melts, $D_{\mathrm{H}}^{\mathrm{met/sil}}$, as a function of temperature, with pressure encoded by color. This figure complements~Figure~\ref{fig:partition_coeff__H} by highlighting the temperature dependence of $D_{\mathrm{H}}^{\mathrm{met/sil}}$ across a range of pressures. Each point corresponds to a unique ${P,T}$ condition. Circles denote thermodynamic-integration \textit{ab initio} molecular dynamics (TI-AIMD) results from this study with $x_{\mathrm{H}}^{\mathrm{sil}} \lesssim 0.2$, and the square marks a two-phase AIMD (2P-AIMD) calculation at $x_{\mathrm{H}}^{\mathrm{sil}} \sim 0.42$. Other symbols represent experimental and computational data from previous work. The horizontal gray line separates siderophile (iron-loving) and lithophile (rock-loving) regimes, underscoring that hydrogen remains strongly siderophilic with only a weak dependence on temperature.
    } 
     \label{fig:partition_coeff__H__T}
\end{figure}

\begin{figure}[h]
\centering
    \includegraphics[width=\textwidth,trim=0 0 0 0,clip]{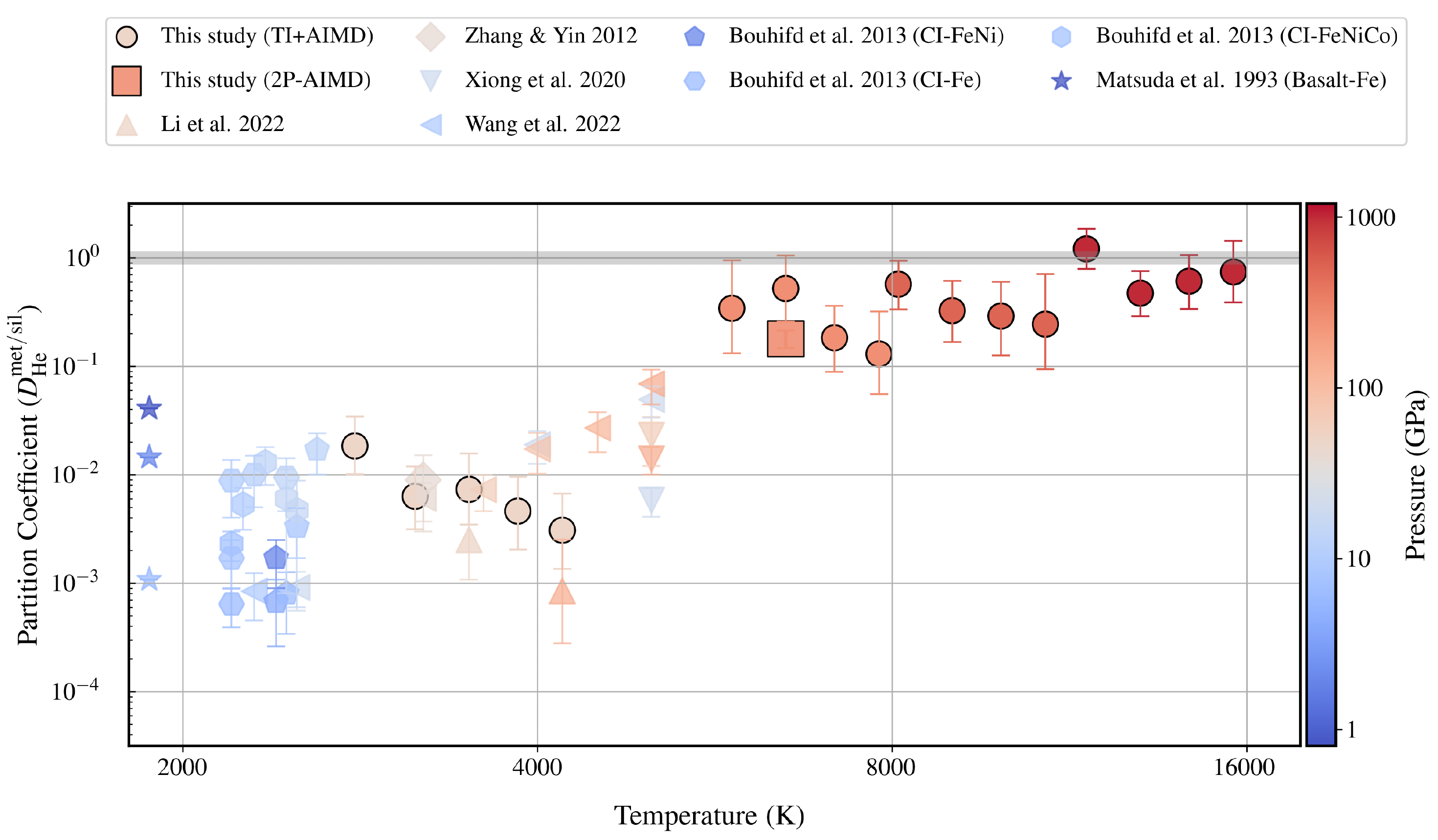}
    \caption{Partition coefficient for helium between silicate and metallic melts, $D_{\mathrm{He}}^{\mathrm{met/sil}}$, as a function of temperature, with pressure encoded by color. This figure complements~Figure~\ref{fig:partition_coeff__He} by emphasizing the temperature dependence of $D_{\mathrm{He}}^{\mathrm{met/sil}}$ across a range of pressures. Each point corresponds to a unique ${P,T}$ condition. Circles and the square represent values calculated in this study at $x_{\mathrm{H}}^{\mathrm{sil}} \lesssim 0.2$ and $x_{\mathrm{H}}^{\mathrm{sil}} \sim 0.49$, respectively, while other symbols show experimental and computational results from previous work. The horizontal gray line separates siderophile (iron-loving) and lithophile (rock-loving) regimes, illustrating helium’s predominantly lithophilic nature with a decreasing sensitivity towards temperature and an increasing affinity towards siderophilicity at higher pressures.
    } 
     \label{fig:partition_coeff__He__T}
\end{figure}

\begin{figure}[h]
\centering
\includegraphics[width=\textwidth,trim=0 0 0 0,clip]{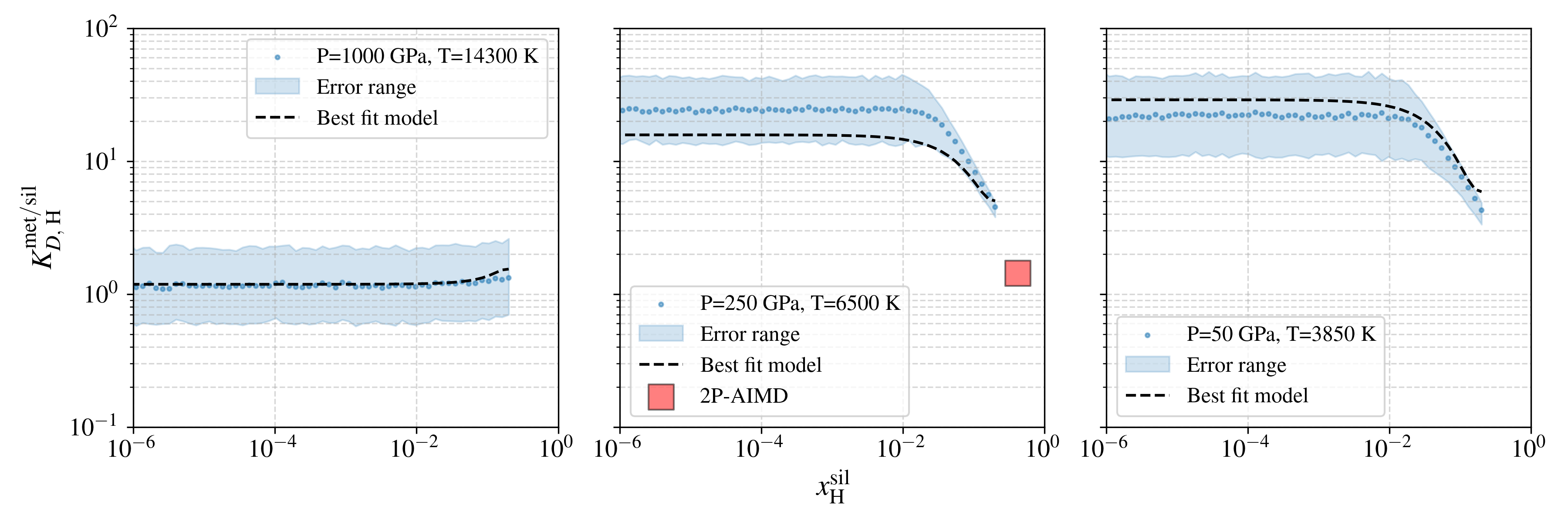}
\caption{Variation of the metal–silicate exchange coefficient $K_{D,\text{H}}^\mathrm{met/sil}$ with pressure, temperature, and composition, along with the global best-fit model. Blue circles denote results from thermodynamic integration combined with ab initio molecular dynamics (TI-AIMD) calculations, and red squares denote results from two-phase simulations (2P-AIMD).
    } 
     \label{fig:array__K_D_chosen__X__H}
\end{figure}

\begin{figure}[h]
\centering
\includegraphics[width=\textwidth,trim=0 0 0 0,clip]{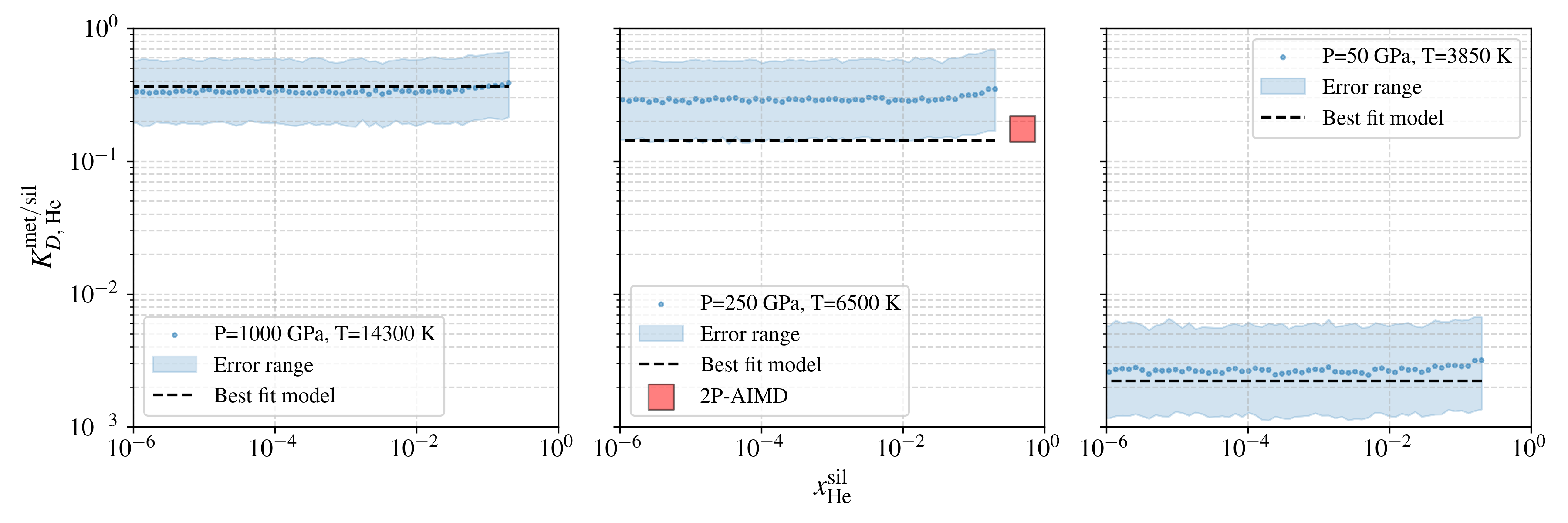}
    \caption{Variation of the metal–silicate exchange coefficient $K_{D,\text{He}}^\mathrm{met/sil}$ with pressure, temperature, and composition, along with the global best-fit model. Blue circles denote results from thermodynamic integration combined with ab initio molecular dynamics (TI-AIMD) calculations, and red squares denote results from two-phase simulations (2P-AIMD).
    } 
     \label{fig:array__K_D_chosen__X__He}
\end{figure}

\begin{figure}[h]
\centering
\includegraphics[width=\textwidth,trim=0 -30 0 230,clip]{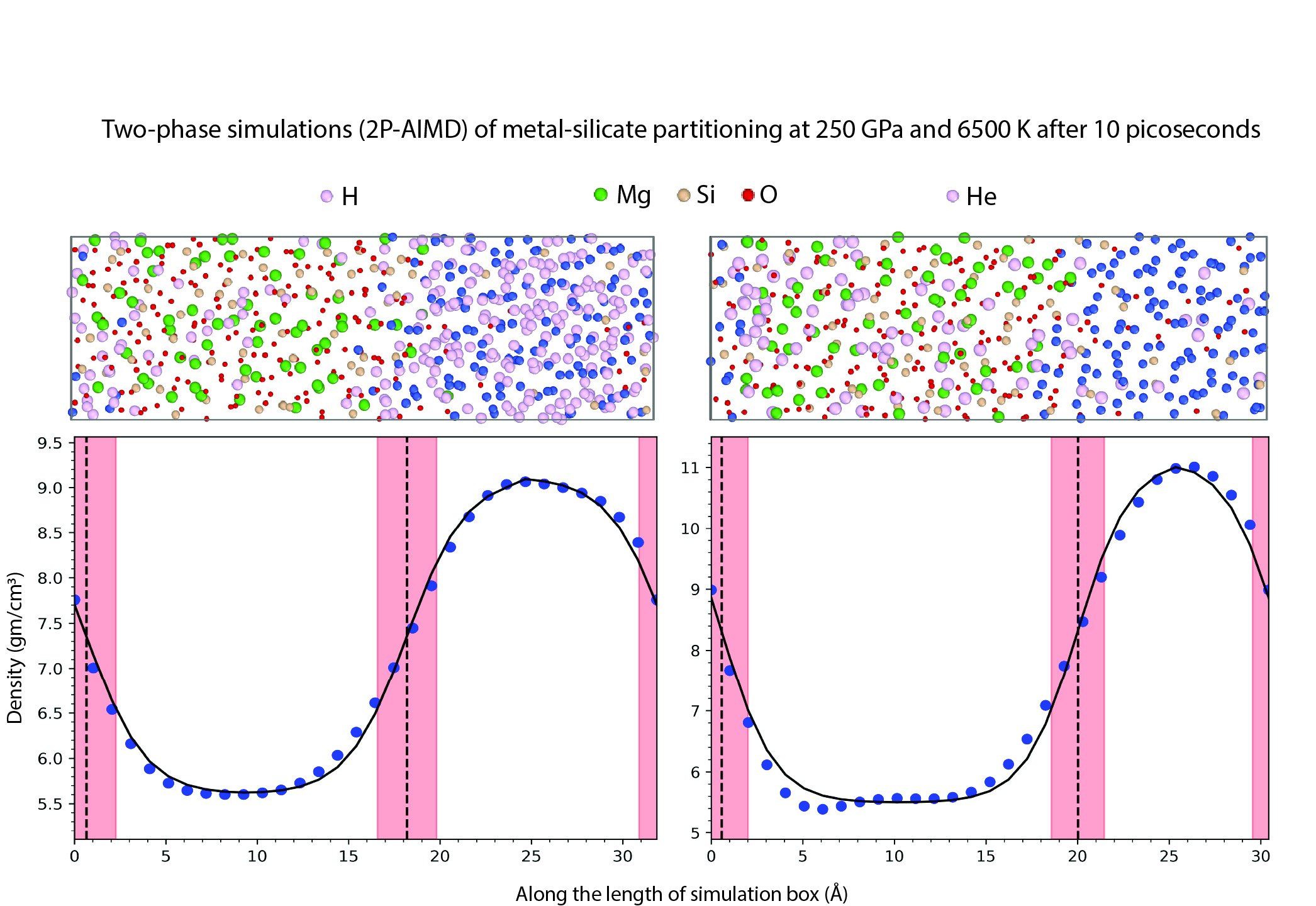}
    \caption{Two-phase ab initio molecular dynamics (2P-AIMD) simulations of H (left) and He (right) metal–silicate partitioning at 250 GPa and 6500 K.
The top panels depict simulation snapshots at 10 ps, while the bottom panels show the corresponding coarse-grained density profiles (filled circles) along the long axis of the simulation cell, together with their best-fit curves (solid lines). The dashed vertical lines mark the Gibbs dividing surfaces that delineate the metallic and silicate phases. H and He abundances are then estimated in these phases over the last 2.5 picoseconds to estimate partition coefficients.
    } 
     \label{fig:2P_AIMD_results}
\end{figure}

\end{document}